\documentclass[twocolumn, aps, prl, superscriptaddress]{revtex4-1}
\usepackage{epsfig}
\usepackage{amsmath}
\usepackage{amssymb}
\usepackage{amsmath}

\usepackage{color}
\newcommand{\mnras}{{Monthly Notices of the Royal Astronomical Society}}
\newcommand{\aap}{{Astronomy \& Astrophysics}}
\newcommand{\araa}{{Ann. Rev. of Astronomy and Astrophysics}}
\newcommand{\solphys}{{Solar Physics}}
\newcommand{\bel}{{\boldsymbol{\mathcal L}}}
\newcommand{\bxi}{{\boldsymbol{\xi}}}
\newcommand{\bnabla}{{\boldsymbol{\nabla}}}
\newcommand{\curl}{{\boldsymbol{\times}}}
\newcommand{\bB}{{\bf B}}
\newcommand{\bv}{{\bf v}}
\newcommand{\ba}{{\bf a}}

\newcommand{\bx}{{\bf x}}
\newcommand{\bs}{{\bf S}}
\newcommand{\blambda}{{\boldsymbol \lambda}}

\begin{document}

\title{Seismic probes of solar interior magnetic structure}

\author{Shravan Hanasoge} \email{hanasoge@princeton.edu}
\affiliation{Department of Geosciences, Princeton University, NJ 08544, USA}
\affiliation{Max-Planck Institut f\"{u}r Sonnensystemforschung, 37191 Kalenburg-Lindau, Germany}
\author{Aaron Birch}
\affiliation{Max-Planck Institut f\"{u}r Sonnensystemforschung, 37191 Kalenburg-Lindau, Germany}
\author{Laurent Gizon}
\affiliation{Max-Planck Institut f\"{u}r Sonnensystemforschung, 37191 Kalenburg-Lindau, Germany}
\affiliation{Georg-August-Universit\"{a}t, Institut f\"{u}r Astrophysik, D-37077 G\"{o}ttingen, Germany}
\author{Jeroen Tromp}
\affiliation{Department of Geosciences, Princeton University, NJ 08544, USA}
\affiliation{Program for Applied and Computational Mathematics, Princeton University, NJ 08544, USA}

\date{March 2012}
                            
\begin{abstract}
Sunspots are prominent manifestations of solar magnetoconvection and
imaging their subsurface structure is an outstanding problem of wide
physical importance. Travel times of seismic waves that propagate through these
structures are typically used as inputs to inversions.
Despite the presence of strongly anisotropic magnetic waveguides, these measurements have always been interpreted in terms of changes to isotropic
wavespeeds and flow-advection related Doppler shifts. 
Here, we employ PDE-constrained optimization to determine the appropriate parameterization
of the structural properties of the magnetic interior.  Seven different wavespeeds fully characterize
helioseismic wave propagation: the isotropic sound speed, a Doppler-shifting flow-advection velocity
and an anisotropic magnetic velocity. 
The structure of magnetic media is sensed by magnetoacoustic slow and fast modes and Alfv\'{e}n waves,
each of which propagates at a different wavespeed.
We show that even in the case of weak magnetic fields, significant errors may be incurred if these anisotropies 
are not accounted for in inversions. Translation invariance is demonstrably lost. These developments render plausible the accurate seismic imaging of 
magnetoconvection in the Sun.
\end{abstract}

\maketitle

Sunspots are substantial deviations from the quiet Sun, with umbral temperatures
dropping by as much as 20\% from ambient conditions. Numerous questions swirl around sunspot physics, such
as understanding their long-time stability (compared to convective turnover timescales) and appreciating
their creation, emergence and eventual death. 
The use of helioseismic waves to probe the structure of sunspots has a long and controversial history
(for a review, see, e.g., \cite{gizon2010}). Inversions for sunspot sub-surface
structure and dynamics (e.g., \cite{Kosovichev1997}) attempt to explain away the observed effects on
seismic waves by an entirely isotropic wavespeed, an approximation that has faced subsequent marginalization (e.g., \cite{gizon_etal_2009})
owing to the widespread recognition of strong anisotropies prevalent in sunspots. Forward modeling
of wave propagation in sunspots has generated a deeper appreciation for measurements and fully realistic non-linear
sunspot evolution calculations (\cite{rempel09}) have proven successful. However, posing an inverse problem
that accounts for these anisotropies remains an outstanding problem of great relevance, towards whose eventual resolution 
this article takes a significant step.

Waves are excited stochastically in the Sun 
due to the action of vigorous near-surface convection. The cross correlation of wavefield velocities measured
at the photosphere of the Sun (by measuring Doppler shifts of absorption lines formed at the photosphere; e.g., the Helioseismic and Magnetic Imager onboard the Solar Dynamics Observatory
 \cite{hmi}) is empirically known to be an ergodic random process (e.g., \cite{gizon_04}).
The associated travel time of a wave between two points on the solar photosphere is estimated by fitting the
cross correlation of the plasma velocities measured at those points. 
We introduce a misfit functional, defined as the $L_2$ norm of the difference between observed ($\tau_i^{\mathrm o}$)
and predicted ($\tau_i^{\mathrm p}$) travel times along a collection of paths $i$:  $\sum_i(\tau_i^{\mathrm o} -\tau_i^{\mathrm p})^2$. 
We then pose the following PDE-constrained optimization problem
\begin{equation}
\chi = \sum_i(\tau_i^{\mathrm o} -\tau_i^{\mathrm p})^2 - \int_\odot d\bx \int d\omega\,\blambda\cdot(\bel\bxi - \bs),\label{con.eq}
\end{equation}
where $\chi$ is the cost function, $\bs$ the wave source, $\omega$ temporal frequency, $\bx$ the spatial coordinate 
and $\blambda(\bx,\omega)$ a vector Lagrange multiplier, the dual to the wave displacement $\bxi(\bx,\omega)$. 
The predicted travel times are linear functionals of wave displacement $\bxi$.
The helioseismic wave operator $\bel$ comprises temporally stationary model properties, which we attempt
to determine.
We reproduce it here in the temporal-frequency domain (e.g., \cite{hanasoge11}),
\begin{eqnarray}
&&\bel\bxi = -\omega^2\rho\bxi -i\omega\rho\Gamma\bxi -2i\omega\rho\bv\cdot\bnabla\bxi\label{waveop} \\
&&- \bnabla(c^2\rho\bnabla\cdot\bxi)- \bnabla(\bxi\cdot\bnabla p) +{\bf g} \bnabla\cdot(\rho\bxi) \nonumber\\
&&  -\frac{1}{4\pi}(\bnabla\curl\bB) \curl [\bnabla\curl(\bxi \curl \bB)] -\frac{1}{4\pi}\{\bnabla\curl[\bnabla\curl(\bxi \curl \bB)]\} \curl \bB,\nonumber
\end{eqnarray}
where the properties of interest are the density $\rho$, sound speed $c$, vector magnetic field $\bB$ and flows $\bv$.
Wave damping is denoted by $\Gamma$ and gravity by ${\bf g}$, where these are held fixed and not considered to be parameters here,
the justifications for which may be found in \cite{hanasoge11}.
Acceleration, wave damping, and Doppler shifting by flow advection are the first two terms of operator~(\ref{waveop}), 
the isotropic wavespeed term and buoyancy terms form the second line and the final two terms are due to the anisotropic Lorentz force.
Background pressure $p$ is constrained by the magneto-hydrostatic (MHS) equilibrium equation 
\begin{equation}
\bnabla p= -\rho{\bf g} + \bnabla\cdot\left(\bB\bB -  \frac{\bB\cdot\bB}{2}\,{\bf I}\right),\label{mhs}
\end{equation}
where ${\bf I}$ is the identity tensor. 
Two scalar thermal parameters, density and sound speed, and two vector quantities, flows and magnetic fields, are the independent variables that we invert for
and pressure is constrained by equation~(\ref{mhs}).

Given wave operator~(\ref{waveop}), we study in detail the variation of the cost function~(\ref{con.eq}) to obtain the 
gradients of the travel-time misfit functional with respect to model parameters (e.g., \cite{tromp10,hanasoge11}). These gradients,
also known as {\it sensitivity kernels} or {\it Fr\'{e}chet derivatives}, are indicators of information in the seismic wavefield and their sensitivity to relevant model
parameters. The computational realization of this method (\cite{tromp10, hanasoge11}) for two-point correlation function measurements, as in the Sun, requires
calculating a predictive so-called {\it forward} wavefield and an {\it adjoint} wavefield that assimilates the misfit.
Sensitivity kernels for various physical quantities (which form the corrector) emerge from a temporal convolution of these two wave fields, allowing us to pose an inverse problem of the form
\begin{equation}
\delta\chi = -\int_\odot d\bx\, K_c\, \delta c + {\bf K}_\bv\cdot\delta\bv + {\bf K}_\bB\cdot\delta\bB + K'_\rho\,\delta\ln\rho,\label{misfit.temp}
\end{equation}
where $K_c, {\bf K}_\bv, {\bf K}_\bB$ are sensitivity kernels for sound speed, flows and magnetic fields, while ${K}'_\rho$ the kernel
for density, termed as an {\it impedance} kernel in geophysics jargon, is sensitive to reflectors (e.g., \cite{zhu09}).
Equation~(\ref{misfit.temp}) states that the measured travel-time shift comprises a sum of volume integrals of these perturbations weighted
by the corresponding finite-frequency wave sensitivities. 
Travel times are very weakly sensitive to density variations but record sharp contrasts in impedance, such as (possibly) the horizontal
boundary of a sunspot. We do not expect to be able to image such reflections since the wavelength of waves that we consider are
large in comparison to possible rapid variations, but retain and compute these kernels in any case (see Figures 2 and 4 of supplemental material).

A critical aspect to setting up an inverse problem is in appreciating the physical variables to which waves are sensitive. 
It is seen that the variation of the operator~(\ref{waveop}) has terms (among others) that contain $\rho\, \delta c^2, \rho\, \delta\bv$ and $\delta\bB$
(see also detailed expressions for kernels in \cite{hanasoge11}). This
immediately tells us that kernels for sound speed and flows are weighted by the density of the model, in contrast to
kernels for the primitive magnetic field. Further, variables $c$ and $\bv$ are forms of wavespeed, which suggests the use of
Alfv\'{e}n velocity, $\ba = \bB/\sqrt{4\pi\rho}$ instead of the primitive $\bB$ field. One may conceive of it as a descriptor of the 
anisotropic wave velocity to which waves are directly sensitive. Straightforward manipulation allows us
to rewrite the kernels as follows 
\begin{equation}
\delta\bB = \delta(\ba\sqrt{4\pi\rho}) = \sqrt{4\pi\rho}\,\delta\ba + \frac{1}{2}\ba\sqrt{4\pi\rho}\, \delta\ln\rho,\label{transform}
\end{equation}
which together with equation~(\ref{misfit.temp}) gives 
\begin{equation}
\sqrt{4\pi\rho}\,{\bf K}_\bB = {\bf K}_\ba\,\,\,\,\,\,\,\,\,\,\,\, K'_\rho \rightarrow K'_\rho + \frac{1}{2}{\bf K}_\ba\cdot\ba,
\end{equation}
thus providing a new expression for variations in the misfit
\begin{equation}
\delta\chi = -\int_\odot d\bx\, K_c\, \delta c + {\bf K}_\bv\cdot\delta\bv + {\bf K}_\ba\cdot\delta\ba + K'_\rho\,\delta\ln\rho.
\end{equation}
We note that the first three terms represent three types of wavespeeds, an isotropic sound speed, an advection related
flow velocity and lastly, an intrinsically anistropic velocity. Although not shown here, weighting the magnetic field kernels by the square-root of density
redistributes incoherent sensitivity from the upper-most atmospheric layers to the photosphere and shallow interior. The transformation for the density kernel in
equation~(\ref{transform}) now contains a contribution from the Alfv\'{e}n velocity, and could in principle be used to
image reflections off sharp velocity contrasts.

The single-scattering first-Born approximation cannot capture the full scope of wave propagation in strong perturbations such as sunspots (i.e., with respect to the quiet Sun; e.g., \cite{gizon06}).
This implies that inversions for the sub-surface structure of sunspots are likely to require an iterative algorithm, since we have to sequentially
refine the predicted travel times, which are nonlinearly related to changes in the model. Thus, in the analysis here, we construct a `sunspot' in MHS equilibrium (Eq.~[\ref{mhs}]) 
and determine sensitivity kernels relative to this model. 

\begin{figure}[t!]
\begin{center}
\includegraphics[width=\linewidth,clip=]{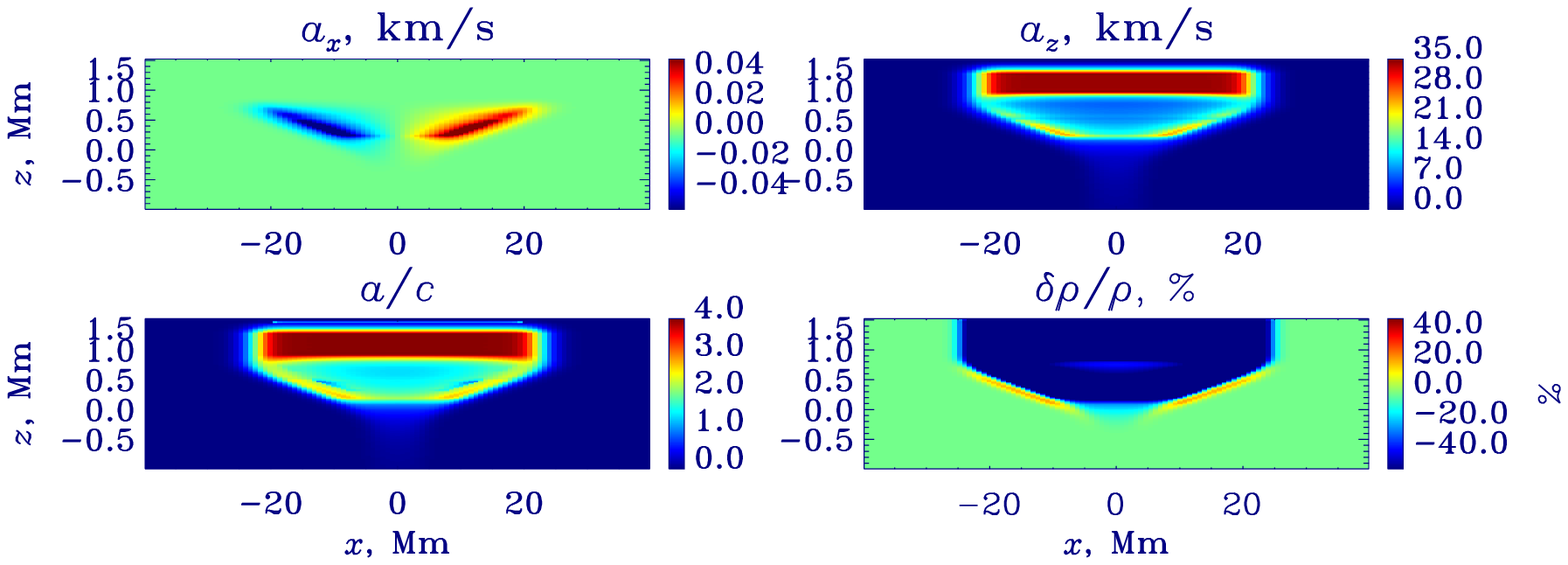}\vspace{-3.5cm}
\end{center}
\caption{Magnetic field configuration in our calculations. Top panels show Aflv\'{e}n speeds
$a_x = B_x/\sqrt{4\pi\rho}$ and $a_z = B_z/\sqrt{4\pi\rho}$, which are signed quantities.
The bottom left panel is the ratio of the absolute Alfv\'{e}n speed to the local sound speed and is seen to be
on the order of 1 at the photosphere. The field is relatively weak with the highest Alfv\'{e}n speed around 35 km/s
and a Wilson depression of 250 km. (for an expanded view, see Figure 1 of the supplemental material)
}
\label{field}
\end{figure}

We introduce a 2-D stream function $\psi(x,z)$ such that the magnetic field is given by $\bB = (-\partial_z\psi,\partial_x\psi)$.
Since ${\bf g} = (0, -g)$, Lorentz forces in the $x$ direction are solely balanced by the pressure gradient in equation~(\ref{mhs}), i.e., $\partial_x p = \partial_x(B_x^2/2 - B_z^2/2) + \partial_z(B_xB_z)$.
From this equation we calculate the pressure distribution required to support this field configuration and then use the $z$ component of equation~(\ref{mhs}) to obtain the associated
density. Generating an MHS state is non trivial since density and pressure decrease exponentially as a function of height above the photosphere; consequently, a large
range of choices for the field configuration results in negative pressures or densities or both. Field configurations with strong horizontal and vertical fields
also require the action of flows to maintain force balance, an aspect we do not consider here because the complexity of such a model renders difficult the interpretation of the attendant kernels. 
We show one example field configuration in Figure~\ref{field}.

A major difficulty in simulating wave propagation through strong magnetic fields is that
Alfv\'{e}n speed $||{\ba}||$ becomes extremely large in the atmospheric layers of the Sun  (due to the exponentially rapidly decreasing density), 
resulting in a very stiff differential equation. Further, wave travel times are very weakly sensitive to the dynamics of these layers because the modes
are trapped below the photosphere. 
A multiplicative prefactor is introduced to control the amplitude of the Lorentz force terms in~(\ref{waveop}), (e.g., \cite{cameron08, rempel09}). However, this method results in
a model that is not seismically reciprocal (e.g., \cite{hanasoge11}), a central requirement in the formal interpretation of helioseismic measurements and the determination
of sensitivity kernels. Here, in order to maintain seismic reciprocity while still saturating the Alfv\'{e}n speed at 40 km/s, we directly multiply the magnetic field by a prefactor. While 
this results in a background field configuration that has a non-zero divergence, we note that small-amplitude oscillations about this field are still divergence free. Further, in the scheme of linear
inversions for magnetic structure, the divergence-free nature of the background field is not a strict requirement but could be considered a regularization term.

We perform linear magneto-hydrodynamic (MHD) wave propagation simulations in Cartesian geometry, 
using the pseudo-spectral code SPARC (\cite{dealias,Hanasoge_couvidat_2008, hanasoge_mag}). Horizontal derivatives are computed using Fast Fourier Transforms,
vertical derivatives are estimated on a non-uniform grid using compact finite differences (\cite{lele92}) and time-stepping is effected through 
the repeated application of an optimized Runge-Kutta scheme (\cite{hu}). Vertical boundaries are lined with absorbent
convolutional perfectly matched layers (\cite{hanasoge_2010}) that are designed to absorb MHD waves as well. 
We implement a phenomenological wave damping term along the lines of the recipe suggested by \cite{schunker11}.
Because we restrict ourselves to a 2-D field configuration in this problem, Aflv\'{e}n waves are disallowed and only magneto-acoustic fast and slow waves propagate.

\begin{figure}[t!]
\begin{center}
\includegraphics[width=\linewidth,clip=]{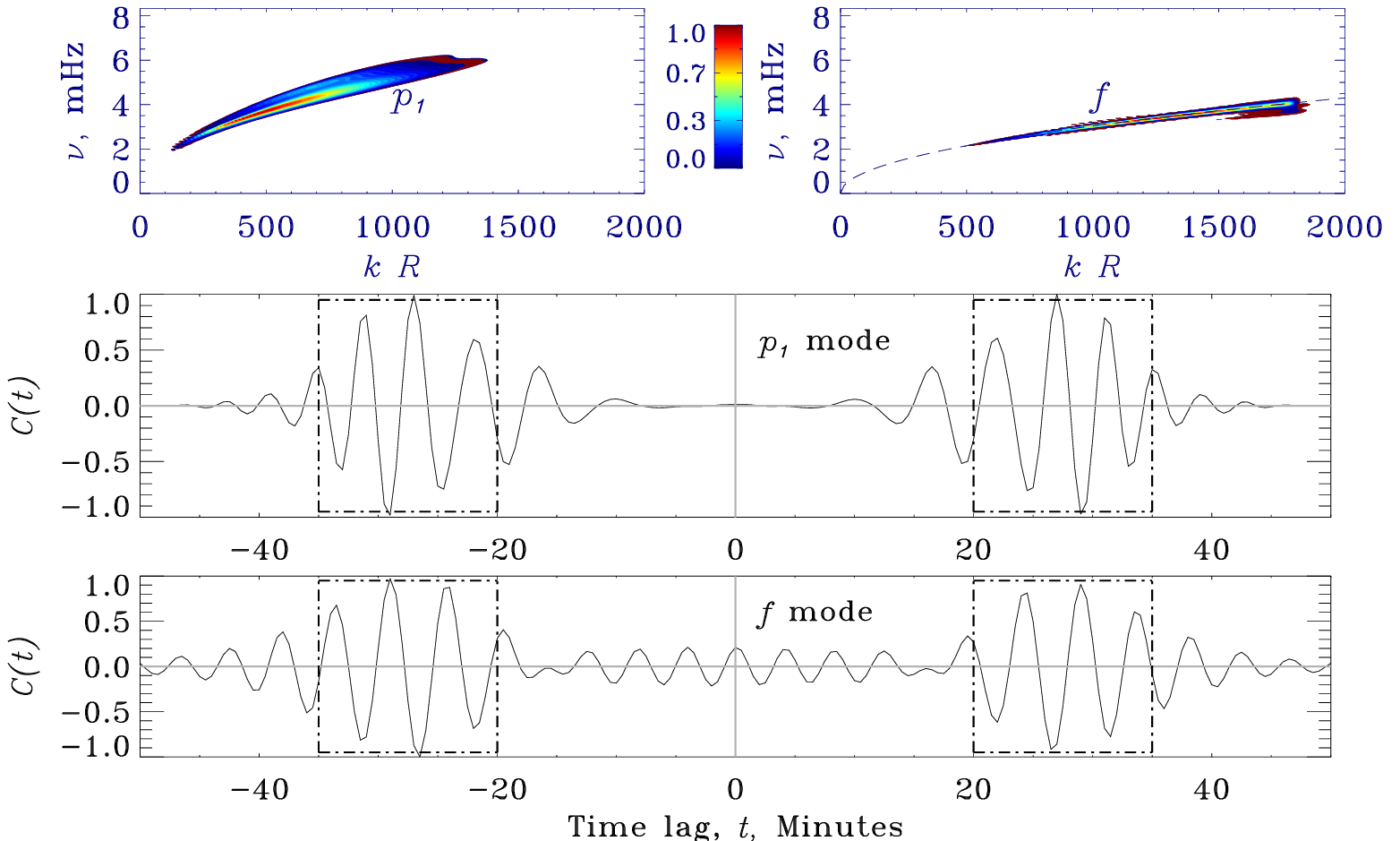}
\end{center}
\caption{Expectation value of the power spectrum of the $p_1$ and $f$ ridge-filtered measurements (top panels). The limit cross correlation ${\mathcal C}(t)$ between a point 15 Mm from the left of the sunspot center to a point 10 Mm on the right of the center is shown for the $p_1$ measurement (middle panel). The $f$-mode cross correlation is between the center of the sunspot and a point 10 Mm to the right (bottom panel). See Figures~\ref{10Mm} and~\ref{25Mm} also. The positive-time branch is sensitive to waves that first arrive at one measurement point and subsequently at the other and vice versa.
The loss of translational variance implies that the absolute locations of the points matter.
The dot-dash boxes indicate the measurement windows. Travel time shifts of waves are obtained by estimating the deviation of the cross correlation from a reference wavelet. {\it Mean} travel
times, defined as the average of the time shifts of oppositely traveling waves, are thought to be largely sensitive to structure. 
{\it Difference} travel times, defined as the difference between the shifts, are considered primarily sensitive to symmetry-breaking flows.
 (for an expanded view, see Figure 2 of the supplemental material)
}
\label{pspec}
\end{figure}

We focus here on the diagnostic ability of the surface $f$ and acoustic $p_1$ modes, so chosen because of their significant sensitivity to surface layers. The measurement consists of
ridge filters applied to isolate these modes. The sunspot is assumed to be located at disk center, implying that the line-of-sight component is co-aligned with the (vertical) $z$ axis. Thus 
the vertical wavefield displacement is used to define the cross correlation measurement.
We show the power spectra and cross correlations in Figure~\ref{pspec}.
We employ the linear travel-time definition (\cite{gizon02,gizon_04}), also used previously by \cite{hanasoge11} to estimate travel time shifts from cross correlations.

\begin{figure}[t!]
\begin{center}
\includegraphics[width=\linewidth,clip=]{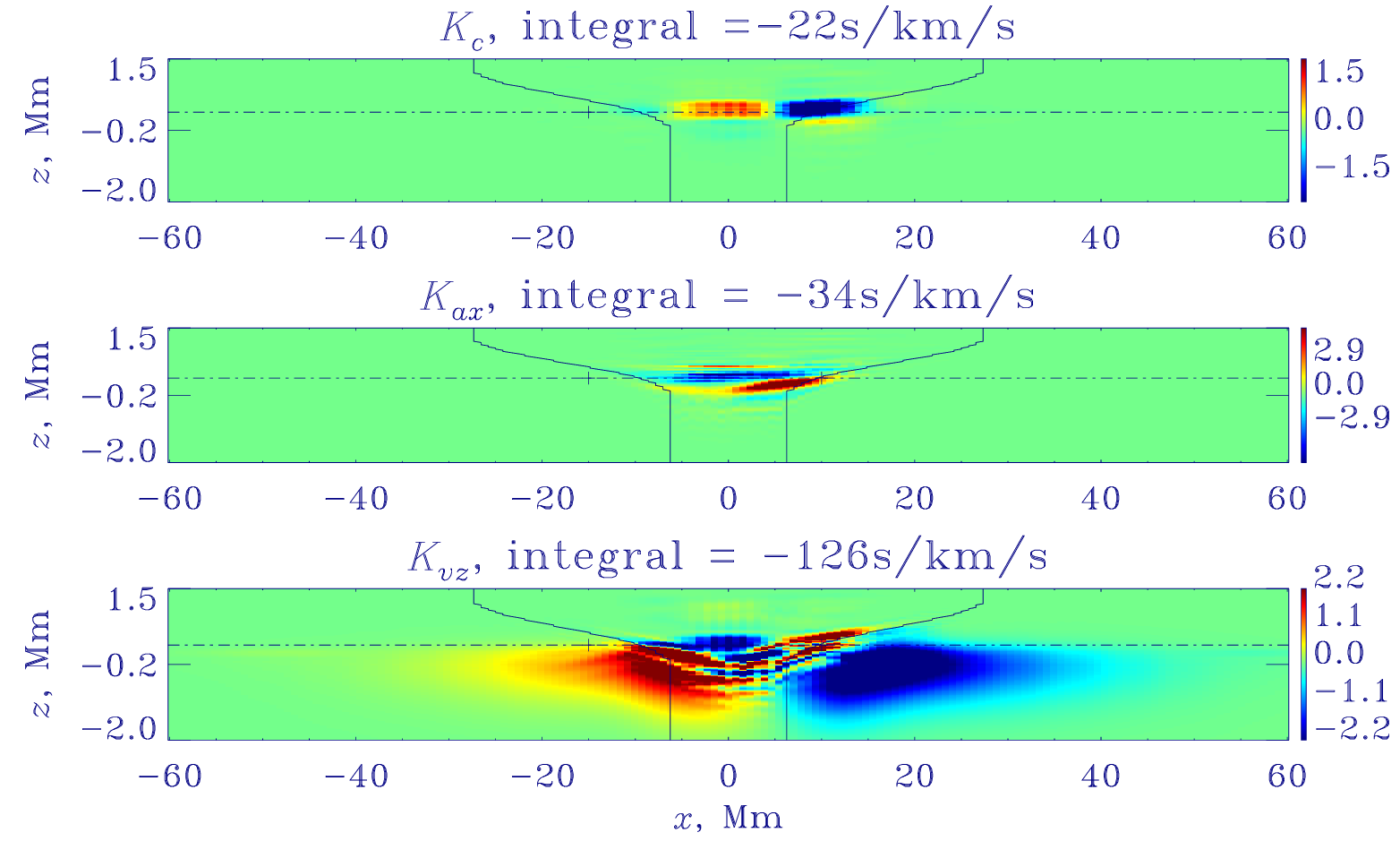}
\end{center}
\caption{$f$-mode (surface) wavespeed kernels for a difference travel-time measurement between a point-pair 10 Mm apart.
Kernels sensitive to isotropic sound speed, Alfv\'{e}n speed $a_x$ and vertical flows $v_z$ are shown.
The boundary of the spot, marked by the solid black line, is much smaller than the horizontal wavelength. 
The horizontal dot-dash line denotes the height at which observations are made in the quiet Sun and the symbols mark the measurement points.
The $f$-mode is seen to be significantly affected by the spot, as seen in the loss in symmetry of the kernels. Signatures of magneto-acoustic slow
and fast modes and hints of conversion to acoustic $p_1$ may be plausibly discerned upon examination. The integrals of the kernels show that
the travel times are significantly affected by the presence of even this relatively weak magnetic field. (for an expanded view, see Figure 3 of the supplemental material)
}
\label{10Mm}
\end{figure}

\begin{figure}[t!]
\begin{center}
\includegraphics[width=\linewidth]{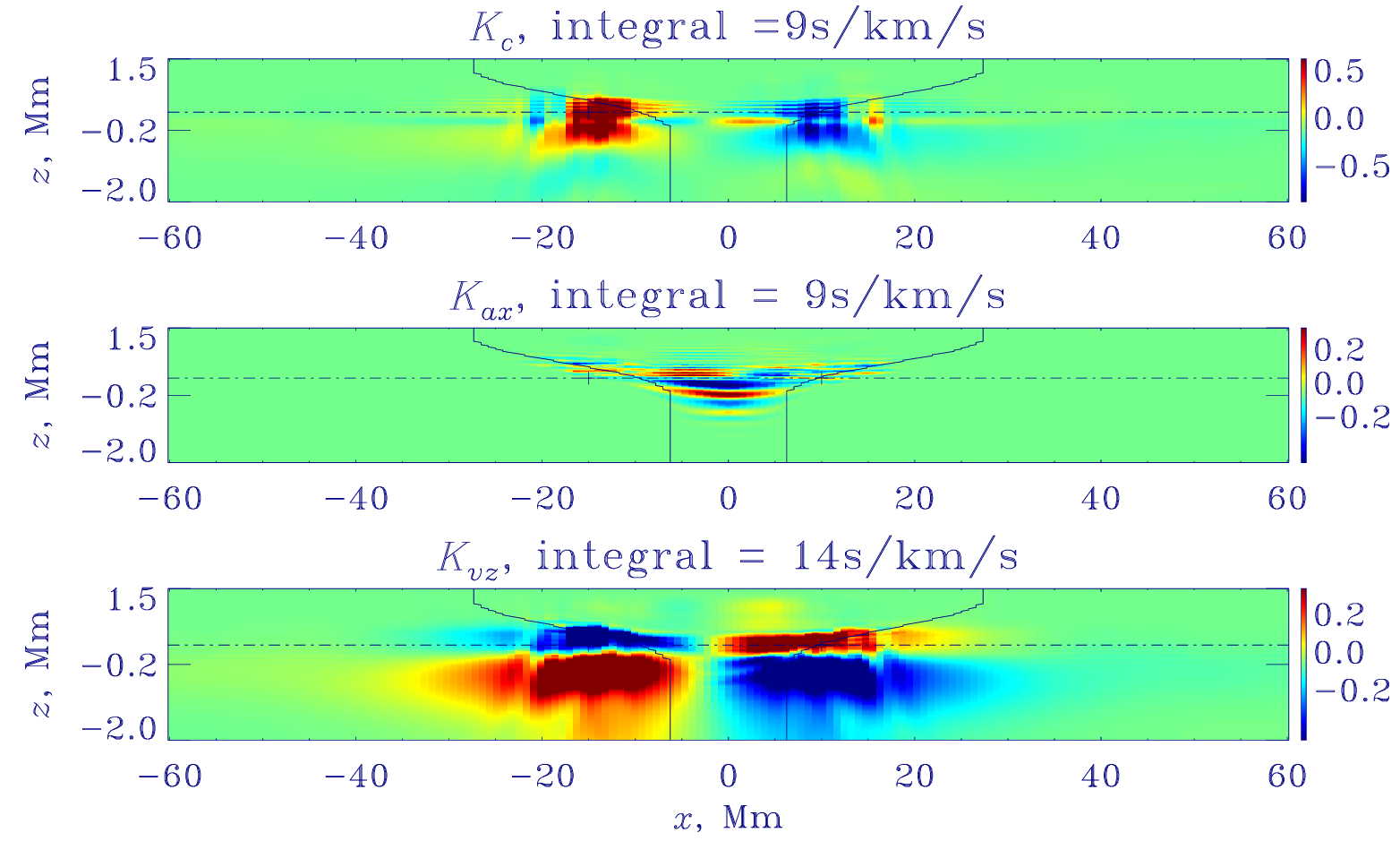}
\end{center}
\caption{$p_1$-mode wavespeed kernels for a difference travel-time measurement between a point-pair 25 Mm apart.
The panels from top to bottom  show kernels sensitive to sound speed (top), Alfv\'{e}n  speeds $a_x$ and vertical flows $v_z$.
The boundary of the spot, marked by the solid black line, is much smaller than the horizontal wavelength.
The horizontal dot-dash line denotes the height at which observations are made in the quiet Sun and the symbols mark the measurement points.
Plausible signatures of slow modes propagating down into the tube may be discerned in the middle panel. (for an expanded view, see Figure 4 of the supplemental material)
}
\label{25Mm}
\end{figure}

Figure~\ref{10Mm}  (see also Figures 5 and 6 in the supplemental material) displays the sensitivity of the surface $f$-mode to the sunspot. 
Because we model waves as finite spatial objects, their sensitivities extend beyond just the ray path.
It can be seen that the effect of
the spot is significant in that the kernels are noticeably asymmetric between the point-pair. The time shifts induced by the magnetic
field are considerable, comparable in magnitude to those induced by flow and thermal perturbations. There are hints of mode conversion
from $f$ to $p_1$ in the difference kernel for sound speed (top), just below the pixel on the right. 

In Figure~\ref{25Mm} (see also Figures 7 and 8 in the supplemental material), we show a set of difference $p_1$-mode kernels for a point pair separated by a distance of 25 Mm respectively. 
Because the magnetic field is relatively weak compared to a sunspot, the acoustic $p_1$ mode, whose energy is focused in the
sub-surface layers, is much less affected by the field than the $f$ mode. Symmetry is nearly completely
restored to the $p_1$ kernels.

The Alfv\'{e}n speed kernels for both $f$ and $p_1$ modes show features of high spatial frequency,
and  contain signatures of fast and slow magneto-acoustic waves. In the umbral regions of the sunspot, waves of high spatial frequency 
are seen to be propagating toward the interior (plausibly slow waves). 



Our computations support the view that inversions for sunspots, especially when using surface modes, are greatly over-simplified
if anisotropic wave speeds are not taken into account.
 The realization of this method has required a number of theoretical and numerical advances, paving the way for seismic imaging of magneto-convection in the solar interior.



{\bf Acknowledgements.}  All calculations were run on the Pleiades
supercomputer at NASA ARC. 
S. M. H. acknowledges support from NASA grant
NNX11AB63G.

\bibliography{kernels}

\end{document}


\title{Seismic probes of solar interior magnetic structure}

\author{Shravan Hanasoge} \email{hanasoge@princeton.edu}
\affiliation{Department of Geosciences, Princeton University, NJ 08544, USA}
\affiliation{Max-Planck Institut f\"{u}r Sonnensystemforschung, 37191 Kalenburg-Lindau, Germany}
\author{Aaron Birch}
\affiliation{Max-Planck Institut f\"{u}r Sonnensystemforschung, 37191 Kalenburg-Lindau, Germany}
\author{Laurent Gizon}
\affiliation{Max-Planck Institut f\"{u}r Sonnensystemforschung, 37191 Kalenburg-Lindau, Germany}
\affiliation{Georg-August-Universit\"{a}t, Institut f\"{u}r Astrophysik, D-37077 G\"{o}ttingen, Germany}
\author{Jeroen Tromp}
\affiliation{Department of Geosciences, Princeton University, NJ 08544, USA}
\affiliation{Program for Applied and Computational Mathematics, Princeton University, NJ 08544, USA}

\date{March 2012}
                            
\begin{abstract}
Supplemental material. Contains expanded figures.
\end{abstract}
\maketitle

Firstly, we show expanded versions of the figures of the main article.

\begin{figure}[t!]
\begin{center}
\includegraphics[width=\linewidth,clip=]{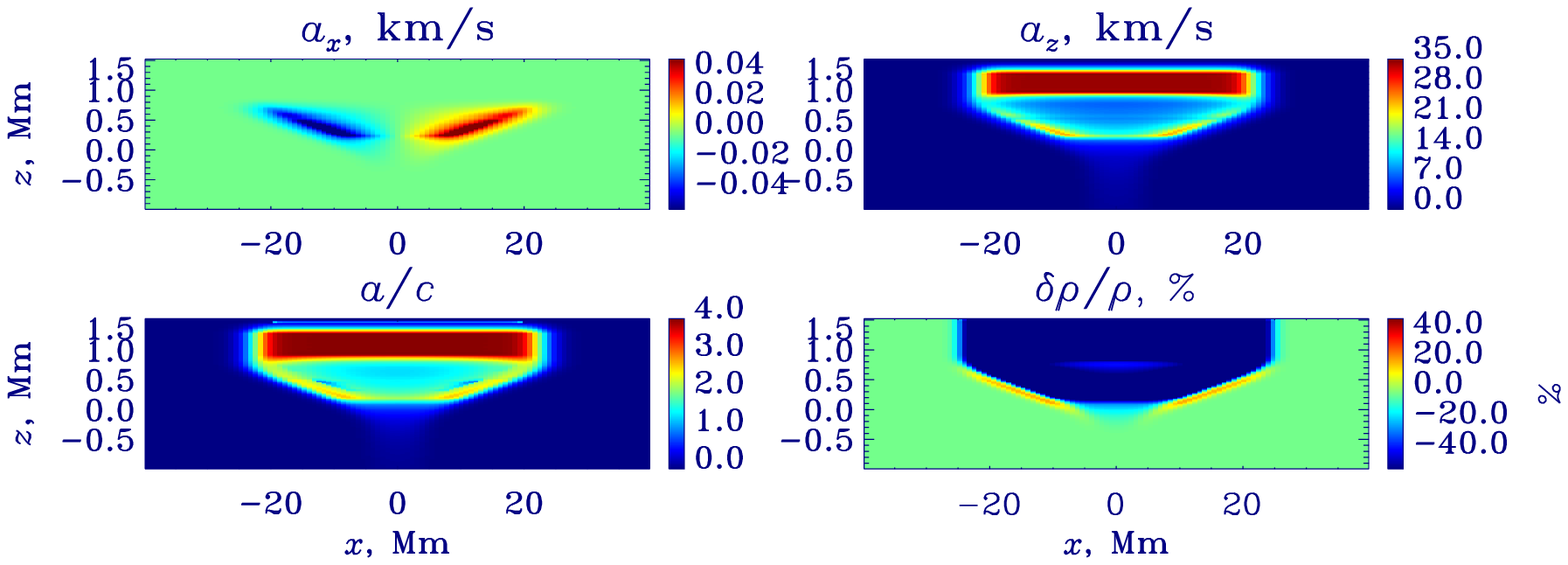}\vspace{-3.5cm}
\end{center}
\caption{Figure 1 of the main article, reproduced: Magnetic field configuration in our calculations. Top panels show Aflv\'{e}n speeds
$a_x = B_x/\sqrt{4\pi\rho}$ and $a_z = B_z/\sqrt{4\pi\rho}$, which are signed quantities.
The bottom left panel is the ratio of the absolute Alfv\'{e}n speed to the local sound speed and is seen to be
on the order of 1 at the photosphere. The field is relatively weak with the highest Alfv\'{e}n speed around 35 km/s
and a Wilson depression of 250 km.
}
\label{field}
\end{figure}

\begin{figure}[t!]
\begin{center}
\includegraphics[width=\linewidth,clip=]{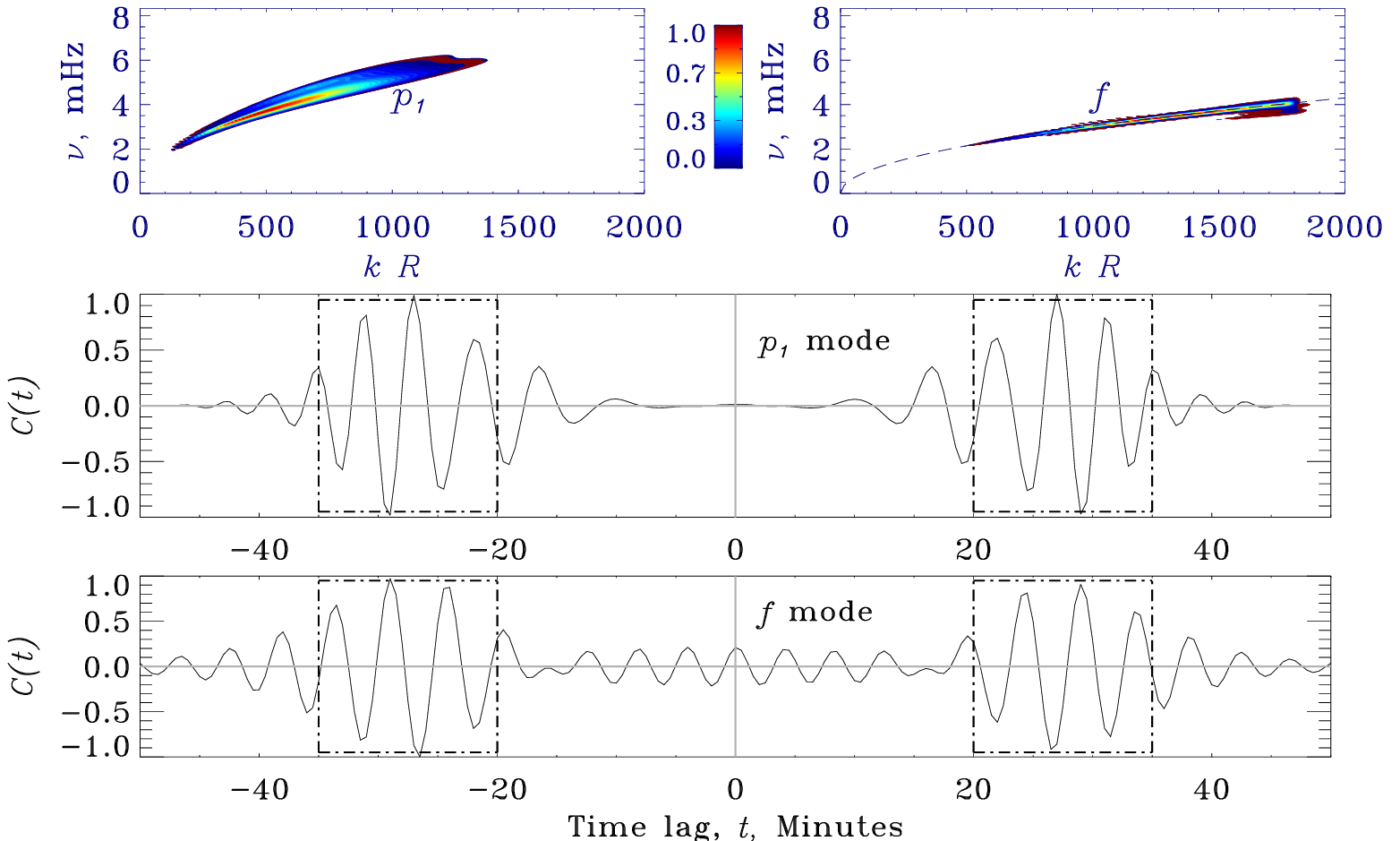}
\end{center}
\caption{Figure 1 of the main article, reproduced: Expectation value of the power spectrum of the $p_1$ and $f$ ridge-filtered measurements (top panels). The limit cross correlation ${\mathcal C}(t)$ between a point 15 Mm from the left of the sunspot center to a point 10 Mm on the right of the center is shown for the $p_1$ measurement (middle panel). The $f$-mode cross correlation is between the center of the sunspot and a point 10 Mm to the right (bottom panel). See Figures~\ref{10Mm} and~\ref{25Mm} also. The positive-time branch is sensitive to waves that first arrive at one measurement point and subsequently at the other and vice versa.
The loss of translational variance implies that the absolute locations of the points matter.
The dot-dash boxes indicate the measurement windows. Travel time shifts of waves are obtained by estimating the deviation of the cross correlation from a reference wavelet. {\it Mean} travel
times, defined as the average of the time shifts of oppositely traveling waves, are thought to be largely sensitive to structure. 
{\it Difference} travel times, defined as the difference between the shifts, are considered primarily sensitive to symmetry-breaking flows.
}
\label{pspec}
\end{figure}

\begin{figure}[t!]
\begin{center}
\includegraphics[width=\linewidth,clip=]{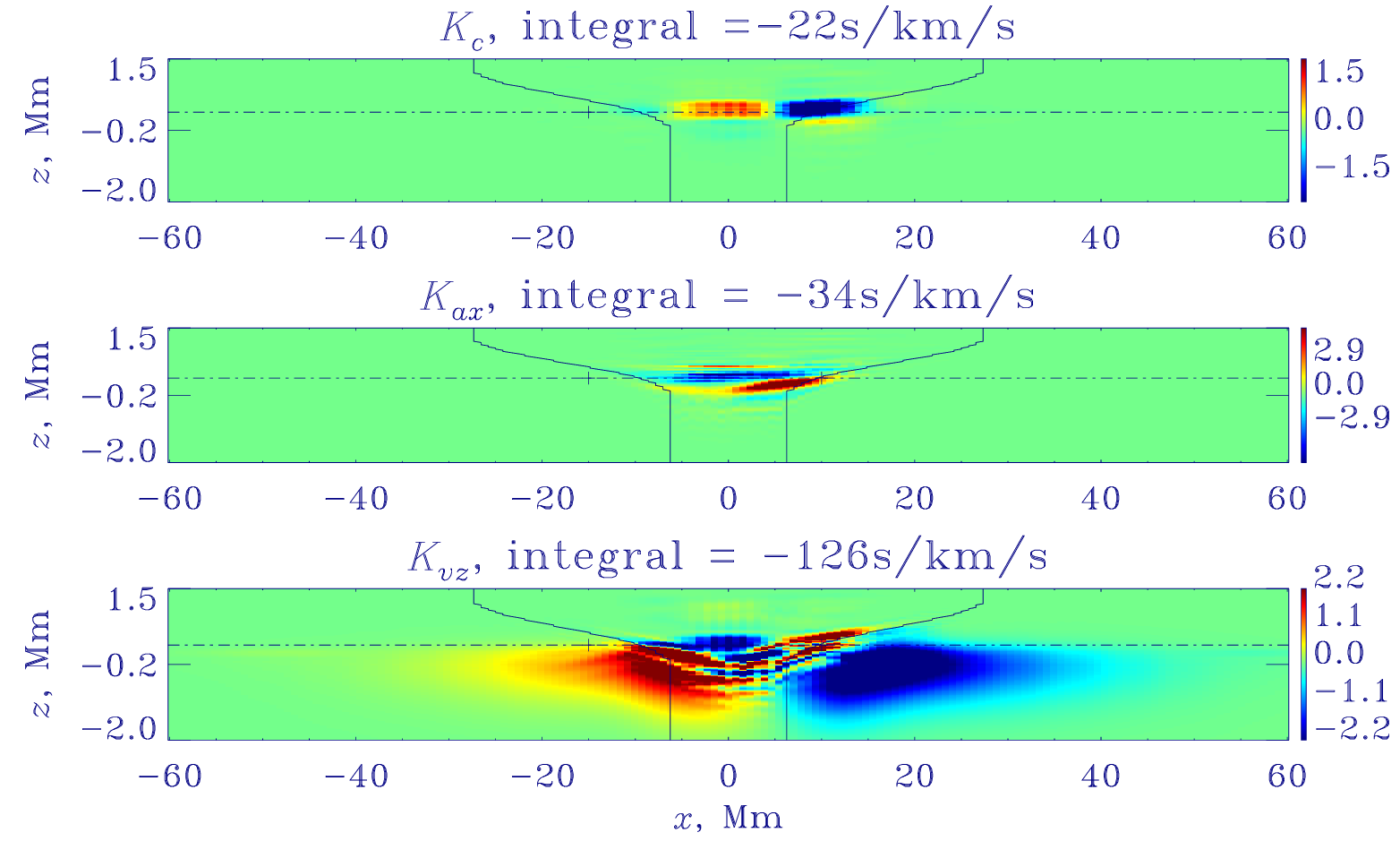}
\end{center}
\caption{Figure 3 of the main article, reproduced: $f$-mode (surface) wavespeed kernels for a difference travel-time measurement between a point-pair separated by a distance of 10 Mm.
Kernels sensitive to isotropic sound speed, magnetic (Alfv\'{e}n) speed $a_x$ and vertical flows $v_z$ are shown.
The boundary of the spot, marked by the solid black line, is much smaller than the horizontal wavelength. 
The horizontal dot-dash line denotes the height at which observations are made in the quiet Sun and the symbols mark the measurement points.
The $f$-mode is seen to be significantly affected by the spot, as seen in the loss in symmetry of the kernels. Signatures of magneto-acoustic slow
and fast modes and hints of conversion to acoustic $p_1$ may be plausibly discerned upon close examination. The integrals of the kernels show that
the travel times are significantly affected by the presence of even this relatively weak magnetic field.
}
\end{figure}

\begin{figure}[t!]
\begin{center}
\includegraphics[width=\linewidth]{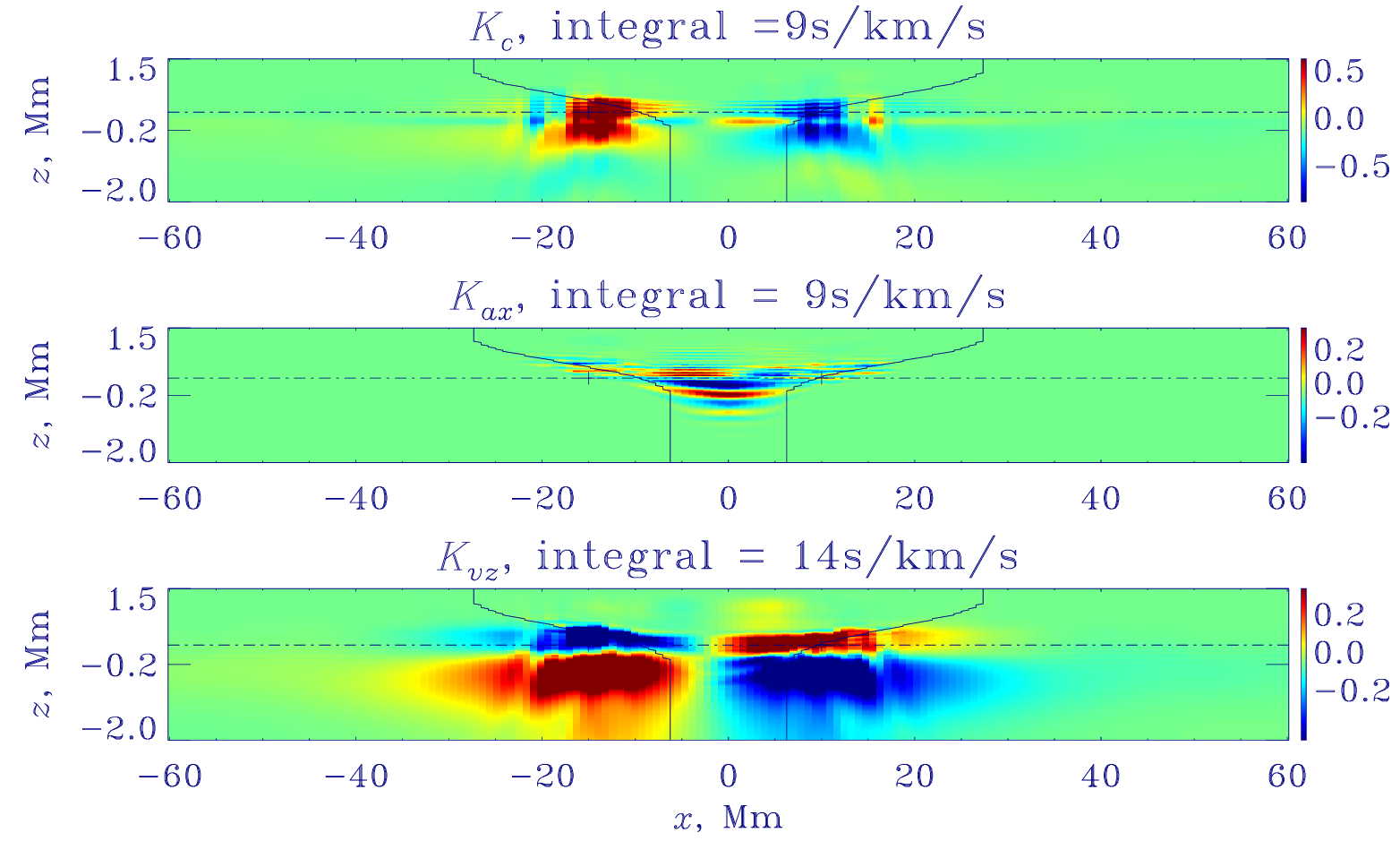}
\end{center}
\caption{Figure 4 of the main article, reproduced:  $p_1$-mode wavespeed kernels for a difference travel-time measurement between a point-pair separated by a distance of 25 Mm.
The panels from top to bottom  show kernels sensitive to sound speed (top), Alfv\'{e}n  speeds $a_x$ and vertical flows $v_z$.
The boundary of the spot, marked by the solid black line, is much smaller than the horizontal wavelength.
The horizontal dot-dash line denotes the height at which observations are made in the quiet Sun and the symbols mark the measurement points.
Plausible signatures of slow modes propagating down into the tube may be discerned in the middle panel.
}
\end{figure}

\begin{figure}[t!]
\begin{center}
\includegraphics[width=\linewidth]{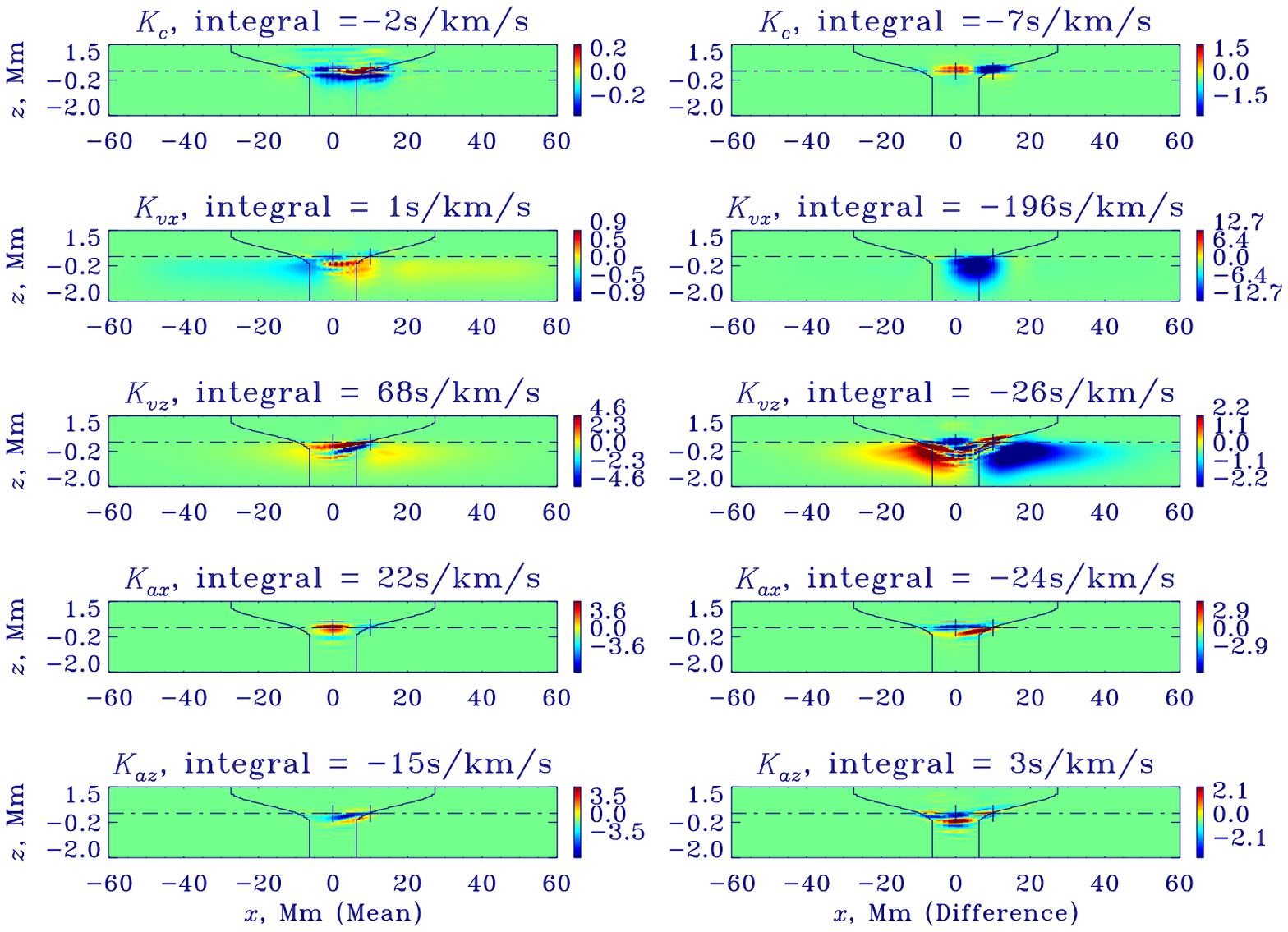}
\end{center}
\caption{$f$-mode (surface) wavespeed kernels for a point-pair separated by a distance of 10 Mm.
The rows display isotropic, advective and anisotropic (Alfv\'{e}n) wavespeeds, in that order.
Kernels on the left column are for the
mean travel time between the point pair whereas the panels on the right column shows kernels for the difference travel time.
The boundary of the spot, marked by the solid black line, is much smaller than the horizontal wavelength. 
The horizontal dot-dash line denotes the height at which observations are made in the quiet Sun and the symbols mark the measurement points.
The $f$-mode is seen to be significantly affected by the spot, as seen in the loss in symmetry of the kernels. Signatures of magneto-acoustic slow
and fast modes and hints of conversion to acoustic $p_1$ may be plausibly discerned upon close examination.
}
\label{10Mm}
\end{figure}

Figure~\ref{10Mm}  displays the sensitivity of the surface $f$-mode to the sunspot. It can be seen that the effect of
the spot is significant in that the kernels are noticeably asymmetric between the point-pair. The time shifts induced by the magnetic
field are considerable, comparable in magnitude those induced by flow and thermal perturbations. There are hints of mode conversion
from $f$ to $p_1$ in the difference kernel for sound speed (top), just below the pixel on the right. Plausible signatures of magneto-acoustic 
slow waves are seen in the difference kernel for $a_z$ and the mean kernel for $v_z$. 

\begin{figure}[t!]
\begin{center}
\includegraphics[width=\linewidth,clip=]{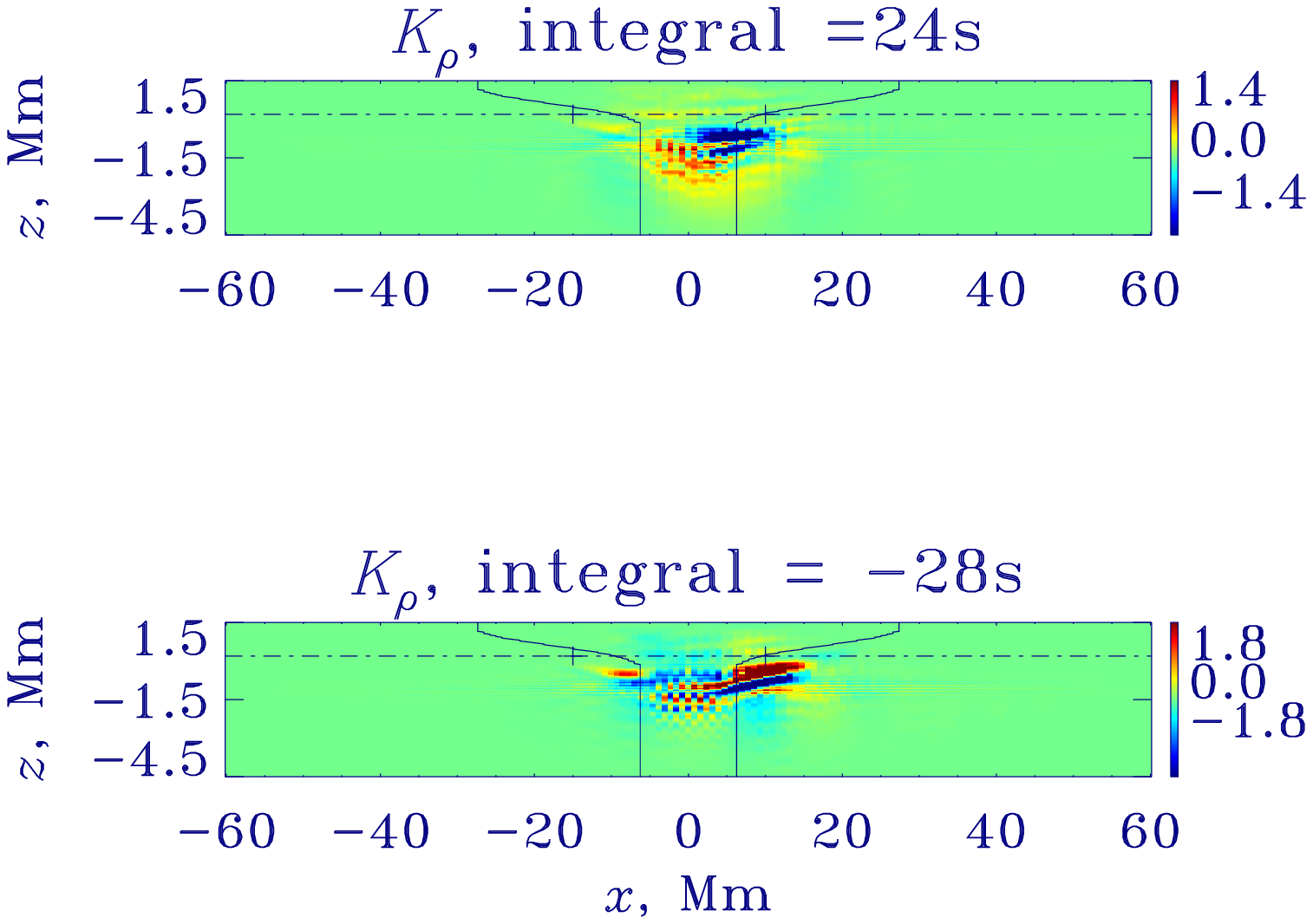}\vspace{-0.5cm}
\end{center}
\caption{$f$-mode (surface) density kernels for mean (upper panel) and difference (lower panel) travel-time measurements between a point-pair separated by a distance of 10 Mm.
The boundary of the spot, marked by the solid black line, is much smaller than the horizontal wavelength. 
The horizontal dot-dash line denotes the height at which observations are made in the quiet Sun and the symbols mark the measurement points.
The $f$-mode is seen to be significantly affected by the spot, as seen in the loss in symmetry of the kernels. The kernel shows significant
complexity, with the difference kernel possessing relatively larger sensitivity to density variations within the spot.}
\label{den10Mm}
\end{figure}

Figure~\ref{den10Mm} shows $f$-mode kernels sensitive to density.

\begin{figure}[t!]
\begin{center}
\includegraphics[width=\linewidth]{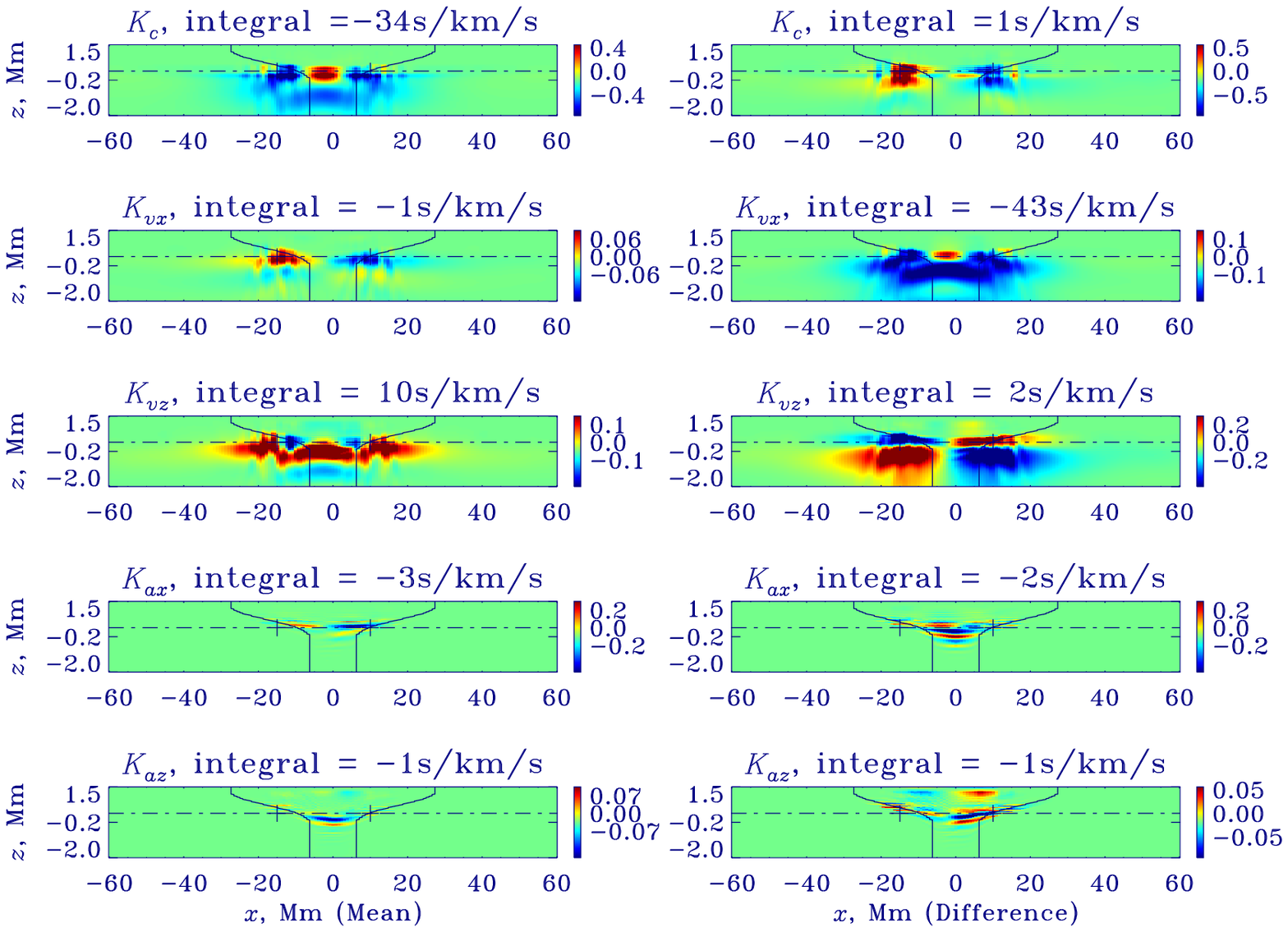}
\end{center}
\caption{$p_1$-mode wavespeed kernels for a point-pair separated by a distance of 25 Mm.
The rows display isotropic, advective and anisotropic (Alfv\'{e}n) wavespeeds, in that order.
 The kernels on the left column are for the
mean travel time between the point pair whereas the panels on the right column shows kernels for the difference travel time.
The boundary of the spot, marked by the solid black line, is much smaller than the horizontal wavelength.
The horizontal dot-dash line denotes the height at which observations are made in the quiet Sun and the symbols mark the measurement points.
}
\label{25Mm}
\end{figure}

In Figure~\ref{25Mm}, we show a set of $p_1$-mode kernels for a point pair separated by a distance of 25 Mm respectively. 
We note right away that the acoustic $p_1$ is less affected by the spot than the $f$ mode.
Advective and Alfv\'{e}n speeds possess similar features, i.e., the $v_x$ and $a_x$ mean travel-time kernels 
both show  lobes of sensitivity near the observation points whereas the $v_z$ and $a_z$ ones show
broad sensitivity between the point-pair. In contrast to both the advective and isotropic wavespeed kernels, 
the Alfv\'{e}n speed kernels only show features of high spatial frequency,
and  contain signatures of fast and slow magneto-acoustic waves. In the umbral regions of the sunspot, waves of high spatial frequency 
are seen to be propagating toward the interior (plausibly slow waves) whereas waves at significant inclination to the the $z$ axis
seem to propagate toward the observation points (plausibly fast waves).

\begin{figure}[t!]
\begin{center}
\includegraphics[width=\linewidth,clip=]{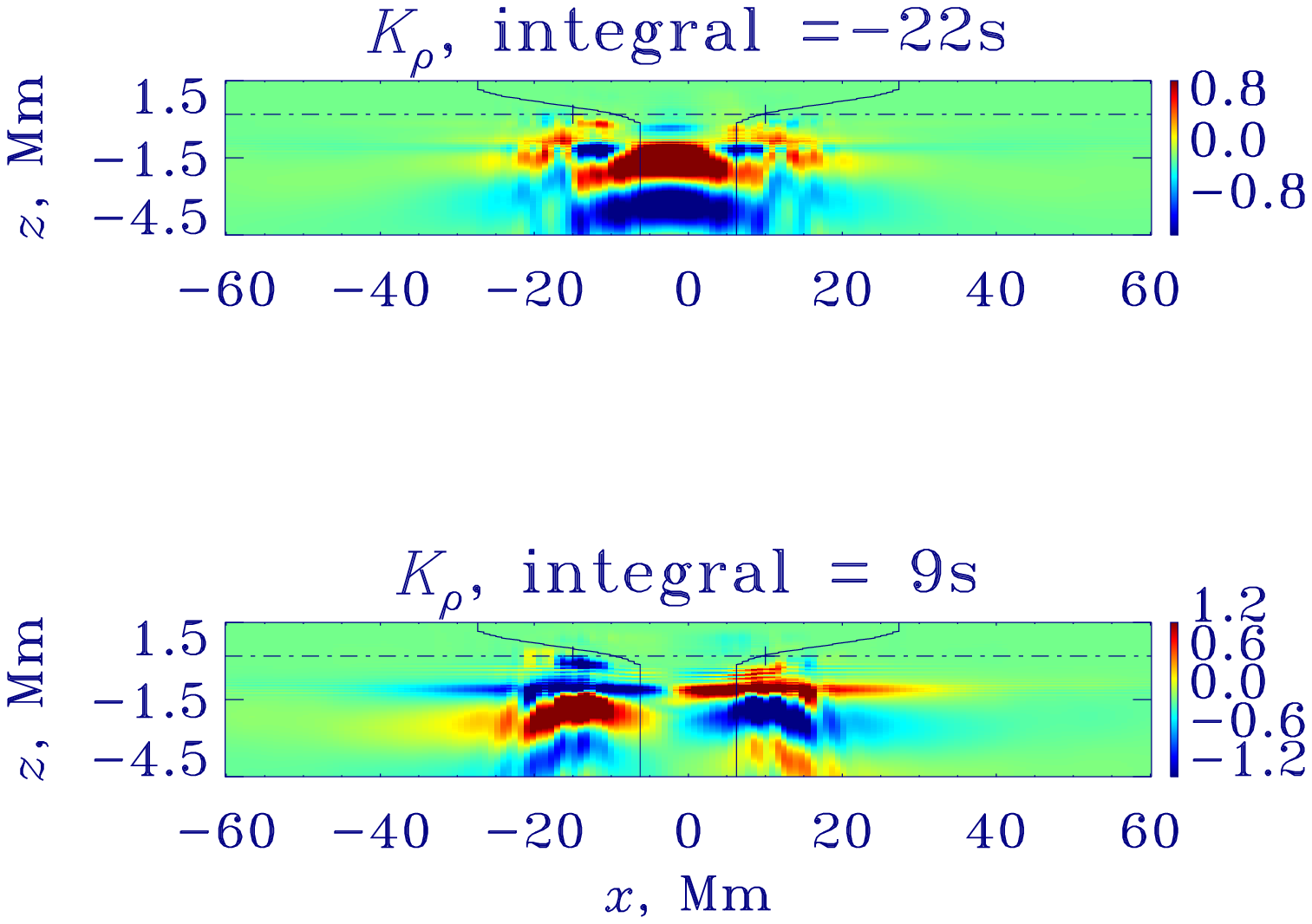}\vspace{-0.5cm}
\end{center}
\caption{Acoustic $p_1$-mode density kernels for mean (upper panel) and difference (lower panel) travel-time measurements between a point-pair separated by a distance of 25 Mm.
The boundary of the spot, marked by the solid black line, is much smaller than the horizontal wavelength. 
The horizontal dot-dash line denotes the height at which observations are made in the quiet Sun and the symbols mark the measurement points.
The $p_1$-mode is seen to be much less affected by the spot than the $f$ mode, as seen in the near symmetry of the kernels. The magnetic spot
is indeed relatively weak and consequently, $p_1$, whose modal energy is focused in the sub-surface layers, is much less sensitive to 
the field.}
\label{den25Mm}
\end{figure}

The sound- and Alfv\'{e}n speed kernels are sensitive 
at different spatial locations to structure, the former being restricted to sub-photospheric layers, the latter showing the 
greatest sensitivity at atmospheric layers where low densities imply very large Alfv\'{e}n speeds. Consequently, a fractional change in 
Alfv\'{e}n speed at these high layers creates travel-time shifts larger than a corresponding change in sound-speed. 
Figure~\ref{den25Mm} shows $p_1$-mode kernels sensitive to density.

